\documentclass[journal,twoside]{IEEEtran}
\usepackage[T1]{fontenc}
\usepackage{hyperref}
\usepackage[square, comma, sort&compress, numbers]{natbib}

\usepackage{graphicx}
\usepackage{subfigure}
\usepackage{multirow}
\usepackage{paralist}
\usepackage{cleveref}
\usepackage{booktabs}
\usepackage{algorithm}
\usepackage{algorithmic}
\usepackage{amsmath}
\usepackage{amsthm}
\usepackage{amssymb}

\usepackage{balance}
\usepackage{microtype}
\usepackage{xcolor}

\usepackage{bm}
\usepackage{array}
\usepackage{color}
\usepackage{colortbl}
\usepackage{makecell}
\definecolor{mygray}{gray}{.9}
\newcolumntype{P}[1]{>{\centering\arraybackslash}p{#1}}
\newcolumntype{M}[1]{>{\centering\arraybackslash}m{#1}}

\newtheorem{Thm}{Theorem}
\newtheorem{Lem}{Lemma}

\newtheorem{Rem}{Remark}

\begin{document}
%
\title{Efficient Soft-Output Gauss-Seidel Data Detector for Massive MIMO Systems}
%
%
%


\author{Chuan~Zhang,~\IEEEmembership{Member,~IEEE},
        Zhizhen~Wu,~\IEEEmembership{Student Member,~IEEE},
        Christoph~Studer,~\IEEEmembership{Senior~Member,~IEEE},
        Zaichen~Zhang,~\IEEEmembership{Senior~Member,~IEEE},
        and~Xiaohu~You,~\IEEEmembership{Fellow,~IEEE},
\thanks{Chuan Zhang, Zhizhen Wu, Zaichen Zhang, and Xiaohu You are with the National Mobile Communications Research Laboratory, Southeast University, Nanjing, China. Christoph Studer is with the School of Electrical and Computer Engineering, Cornell University, Ithaca 14853, NY. Email: chzhang@seu.edu.cn.}
\thanks{This paper was presented in part at IEEE International Symposium on Circuits and Systems (ISCAS), Montreal, Canada, 2016 \cite{Me2016iscas}. Chuan Zhang and Zhizhen Wu have the same contribution to this work. \emph{(Corresponding author: Chuan Zhang.)}}
}

%
%

\markboth{}%
{C. Zhang \MakeLowercase{\textit{et al.}}: Efficient Soft-Output Gauss-Seidel Data Detector for Massive MIMO Systems}
%



\maketitle

\begin{abstract}
For massive multiple-input multiple-output (MIMO) systems, linear minimum mean-square error (MMSE) detection has been shown to achieve near-optimal performance but suffers from excessively high complexity due to the large-scale matrix inversion. Being matrix inversion free, detection algorithms based on the \emph{Gauss}-\emph{Seidel} (GS) method have been proved more efficient than conventional \emph{Neumann} series expansion (NSE) based ones. In this paper, an efficient GS-based soft-output data detector for massive MIMO and a corresponding VLSI architecture are proposed. To accelerate the convergence of the GS method, a new initial solution is proposed. Several optimizations on the VLSI architecture level are proposed to further reduce the processing latency and area. Our reference implementation results on a Xilinx Virtex-7 XC7VX690T FPGA for a 128 base-station antenna and 8 user massive MIMO system show that our GS-based data detector achieves a throughput of 732\,Mb/s with close-to-MMSE error-rate performance. Our implementation results demonstrate that the proposed solution has advantages over existing designs in terms of complexity and efficiency, especially under challenging propagation conditions.
\end{abstract}


%
\IEEEpeerreviewmaketitle
\section{Introduction}
\IEEEPARstart{T}{he} transmission of multiple data streams concurrently and in the same frequency band, which is known as multiple-input multiple-output (MIMO) technology, enables higher data rates compared to traditional single-input single-output (SISO) wireless communication systems~\cite{Foschini1998}. Due to the combined effect of the ever-growing mobile data traffic and the scarcity of favorable radio spectrum in the low-loss frequency range, massive MIMO is widely believed to be a core technology for upcoming fifth generation (5G) wireless communication systems \cite{Rusek2013}. Massive MIMO  promises higher data rates, improved spectral efficiency, better link reliability, and coverage over small-scale MIMO systems \cite{Mietzner2009,Gesbert2010,Andrews2014,Lu2014,Han2011,Ngo2013}. Compared to SISO systems, multiple interfering messages/symbols are transmitted concurrently in MIMO systems and expected to be separated at the receiver side with the contamination of noise or interference \cite{Yang2015}. In the case of massive MIMO, however, this data detection operation entails excessively high computational complexity with optimal methods as the number of base-station and user antennas increases significantly.

\subsection{Detection in Massive MIMO Systems}
The optimal data detection problem in MIMO systems is non-deterministic
polynomial-time hard (NP-hard) \cite{van1981another,Verdu1989,Micciancio2001}. Hence, existing algorithms that aim at solving this problem optimally, e.g., algorithms based on the optimal maximum-likelihood (ML) criterion \cite{ML:verdu1986minimum} or the maximum \emph{a posteriori} (MAP) criterion \cite{Moher}, inevitable require excessively high complexity as the number of decision variables increases with the number of transmitted data streams. Though the hardware computing capability has evolved significantly over recent decades, efficient hardware implementations for optimal data detection remains challenging. On the other hand, the new emerged application scenarios such as  massive machine type communications (mMTC) and ultra-reliable low latency communications (URLLC) are expecting massive MIMO detectors of low complexity, latency, and area, which implies that even data detectors that have modest-complexity would be unacceptable due to the stringent power and area constraints \cite{Yang2015}. Therefore, near-optimal, low-complexity, and high-speed massive MIMO detectors are highly desired to bridge the gap between algorithms and hardware implementations.

One possible solution {to perform optimal data detection}  are non-linear data detection algorithms with reduced complexity, such as the sphere decoder (SD) \cite{Wong2002,Hochwald2003,Studer2008} and tabu search (TS)-based data detectors \cite{Barbero2008,Srinidhi2011}. Admittedly, such solvers worked perfectly well for traditional, small-scale MIMO systems. However, for massive MIMO systems with tens to hundreds of antennas and higher-order modulation schemes \cite{Jalden2005,Datta2012}, such methods result in prohibitively high computation and implementation complexity \cite{Verdu1989}.

Alternatively, one can resort to linear data detection algorithms, such as zero-forcing (ZF) or minimum mean-square error (MMSE)-based equalization, to tradeoff performance versus complexity \cite{Rusek2013}. Unfortunately, both of these methods require a large-dimensional  matrix inversion. Exact inversion algorithms of an $N \times N$ matrix, using, for example, QR-\emph{Gram Schmidt} \cite{Singh2007}, \emph{Gauss-Jordan} \cite{Arias-Garcia2011}, or \emph{Cholesky} decomposition \cite{Studer2011}, entail high complexity of  $\mathcal{O}(N^3)$. By exploiting the channel hardening property of wireless channels in massive MIMO systems, reference \cite{Wu2014nse} proposed a \emph{Neumann} series expansion (NSE)  that reduces the complexity of matrix inversion. However, this algorithm still suffers from a complexity of $\mathcal{O}(N^3)$ when the NSE length $K \geq 2$ (even higher than the exact inverse when $K \geq 4$) \cite{Wu2014nse}. To achieve a lower complexity of $\mathcal{O}(N^2)$, a soft-output detection based on \emph{Gauss}-\emph{Seidel} (GS) method was proposed in \cite{Dai2015} recently. But the existing GS-based detectors still exhibit slow convergence rates and relatively high hardware efficiency.

\subsection{Contributions}
In this paper, an efficient GS-based soft-output data detection algorithm and a corresponding VLSI architecture are proposed. Based on the fact that the MMSE filtering matrix is diagonally dominant for massive MIMO systems, a $2$-term NSE is employed to generate the initial solution of the GS method, which effectively accelerates the convergence, especially under challenging propagation environments, {such as MIMO systems with a large system loading factor or correlated channels}.
We provide a VLSI architecture along with numerous architecture- and  hardware-level optimizations. {Our implementation results demonstrate that the proposed detector achieves a throughput of 732\,Mb/s with close-to-MMSE error-rate performance, outperforming state-of-the-art designs in terms of complexity and efficiency.}
With the aid of the proposed efficient architecture, the GS-based detector is flexible and suitable to meet various system requirements.

\subsection{Outline of the Paper}

The reminder of the paper is organized as follows. Section~\ref{sec:pre}  reviews the prerequisites. Section \ref{sec:algo} proposes the GS-based soft-output data detection algorithm for massive MIMO systems and provides a complexity analysis and error-rate performance comparison. Section \ref{sec:arch} details the VLSI architecture with several optimizations. Section \ref{sec:imple} provides reference FPGA implementation results and a comparison with existing designs. Section \ref{sec:con} concludes the paper.

\subsection{Notation}

Lower- and upper-case boldface letters stand for column vectors and matrices, respectively. The entry in the $i$th row and $j$th column of a matrix $\mathbf{A}$ is represented by $A_{ij}$; the $k$th entry of a vector $\mathbf{a}$ is represented by $a_k$.
The operations $(\cdot)^H$, $(\cdot)^T$, $(\cdot)^{-1}$ denote conjugate transpose, transpose and inverse respectively; $|\cdot|$ and $\lVert\cdot\rVert$ denote the absolute value operator and the Euclidean norm respectively.
$\mathbb{E}\{\cdot\}$ stands for expectation.
$\mathbf{I}_K$ denotes the $K \times K$ identity matrix.
The superscript $(\cdot)^{(k)}$ indicates the $k$th iteration of iterative methods.

\section{Prerequisites}\label{sec:pre}

\subsection{System Model}
We consider a multi-user MIMO uplink system ith $N_t$ users and $N_r$ receive antennas at the base station (BS). In the case of massive MIMO, we assume that $N_r \gg N_t$. Spatial multiplexing is employed and each user is equipped with a single antenna\footnote{The model can be extended to the case of multiple spatial streams per user.}. For each user, the information bits $\mathbf{b}$ are encoded into the coded bit-stream $\mathbf{x}$ and then mapped to the transmit vector $\mathbf{s}=[s_1,\ldots,s_{N_t}] \in \Omega^{N_t}$, where $\Omega$ corresponds to the $2^B$-QAM constellation using Gray labelling. Therefore, each transmit vector is associated with $N_tB$ information bits and {$x_{ib}$}  denotes the $b$th bit of the $i$th entry of $\mathbf{s}$. The transmit vector $\mathbf{s}$ is transmitted over a wireless MIMO channel modeled as
\begin{equation}
\mathbf{y}=\mathbf{Hs}+\mathbf{n},
\end{equation}
where $\mathbf{y}=[y_1,\ldots,y_{N_r}]^T \in \mathbb{C}^{N_r}$ corresponds to the received vector at the BS, $\mathbf{H} \in \mathbb{C}^{N_r \times N_t}$ is the channel matrix, and $\mathbf{n} \in \mathbb{C}^{N_r}$ stands for independent identically distributed (i.i.d.) Gaussian noise with mean zero and variance $N_0$ per entry. In the following, the channel matrix $\mathbf{H}$ and the noise variance $N_0$ are assumed to be perfectly known at the receiver. The transmit symbol variance is normalized as $E_s=\mathbb{E}[|s_i|^2]=1$. In what follows, we employ the average SNR per receive antenna defined as $\mathrm{SNR}=N_tE_s/N_0$.

\subsection{Soft-Output MMSE Data Detection}
\label{subsec:sommse}
The task of the BS is to compute log-likelihood ratio (LLR) values for the coded
bits given $\mathbf{H}$ and $\mathbf{y}$ with a soft-output data detection algorithm. Since MMSE algorithms have been proven to be near-optimal for massive MIMO uplink with low complexity \cite{Wu2014nse}, we consider the typical linear MMSE detection, which can be written as
\begin{equation}
\label{eq:detection}
\hat{\mathbf{s}}= \mathbf{W}^{-1}\mathbf{y}^\mathrm{MF},
\end{equation}
where $\mathbf{y}^\mathrm{MF}=\mathbf{H}^H\mathbf{y}$ is the matched-filter output and $\mathbf{W}=\mathbf{G}+N_0\mathbf{I}_{N_t}$ is the regularized Gram matrix with the Gram matrix $\mathbf{G}=\mathbf{H}^H\mathbf{H}$.
In order to obtain the LLR values, \eqref{eq:detection} can be rewritten as
\begin{equation}
\hat{\mathbf{s}}=\mathbf{U}\mathbf{s}+\mathbf{W}^{-1}\mathbf{H}^H\mathbf{n},
\end{equation}
where $\mathbf{U}=\mathbf{W}^{-1}\mathbf{G}$ is the equivalent channel matrix. Then the equalized symbol of the $i$th user can be written as $\hat{s}_i=\mu_is_i+t_i$, where $\mu_i=\mathbf{U}_{ii}$ denotes the effective channel gain and $t_i$ denotes noise-plus-interference (NPI).
The $a \ posteriori$ LLRs for each bit of $\mathbf{x}$ can be computed by
\begin{equation}
L_{ib} = \ln \left( \frac{\Pr[x_{ib}=1|\mathbf{y}]}{\Pr[x_{ib}=0|\mathbf{y}]} \right),\; i=1,\ldots,N_t, b=1,\ldots,B.
\end{equation}

Entailing only a small loss for high-order modulation schemes, the approximation of LLR computation proposed in \cite{Collings2004} is hardware-friendly for Gray mappings. Therefore, an efficient approach to compute the extrinsic LLRs of the detector is given by
\begin{equation}
\label{eq:apprLLR}
\hat{L}_{ib}=\rho_i\left(\min_{a \in \Omega_b^0}\left|
z_i-a\right|^2-\min_{a' \in \Omega_b^1}\left|
z_i-a'\right|^2\right)=\rho_i\lambda_b(z_i),
\end{equation}
where $z_i=\hat{s}_i/\mu_i$ and $\rho_i=\mu_i^2/\nu^2_i=\mu_i / (1-\mu_i)$ represents post-equalization signal-to-interference-and-noise-ratio (SINR), $\Omega_b^0$ and $\Omega_b^1$ denote subsets of $\Omega$ for which the $b$-th bit is 0 and 1, respectively.   {As summarized in \cite{Collings2004}, the function $\lambda_b(z_i)$ can be computed efficiently for Gray mappings.}  Note that the computation of the LLRs is the same for both real parts and imaginary parts.

\section{Low-Complexity Signal Detection Algorithm}\label{sec:algo}
In this section, a low-complexity soft-output detection algorithm named  {improved GS detection (IGS)} is proposed for massive MIMO uplink. Algorithm \ref{alg:igs} provides a summary of the method.

\subsection{Signal Detection with Gauss-Seidel Method}\label{subsec:gs}

Computing the inverse of the regularized Gram matrix $\mathbf{W}^{-1}$ requires a computational complexity of $\mathcal{O}(N_t^3)$ using hardware-friendly matrix inverse approaches, which is prohibitive for massive MIMO. Therefore, iterative methods with low complexity have been exploited in massive MIMO detection. The GS method is used to solve the $N$-dimension linear equation $\mathbf{Ax=b}$, where $\mathbf{A}$ is the $N \times N$ coefficient matrix, $\mathbf{x}$ is the solution vector, and $\mathbf{b}$ is the measurement vector. In what follows, the reason why the GS method can be utilized to solve massive MIMO detection problem is explained.
\begin{Lem}\label{lem:hpd}
For massive MIMO systems, the columns of channel matrix $\mathbf{H}$ are asymptotically orthogonal, and the regularized Gram matrix $\mathbf{W}$ is Hermitian positive definite with probability one.
\end{Lem}
\begin{IEEEproof}
Please refer to \cite{Rusek2013} and \cite{Gao2014} for the proof.
\end{IEEEproof}


\begin{Thm}\label{thm:gs}
The GS method converges for any initial solution if $\mathbf{A}$ is Hermitian positive definite.
\end{Thm}
\begin{IEEEproof}
Refer to \cite[Chapter~4]{Saad2003iterative} for a proof.
\end{IEEEproof}


Specifically,  signal detection in massive MIMO systems using the GS method is carried out by the following steps:

\emph{Step 1:} Decompose the regularized Gram matrix $\mathbf{W}$ as
\begin{equation}
\mathbf{W}=\mathbf{D}+\mathbf{L}+\mathbf{L}^H,
\end{equation}
where $\mathbf{D}$, $\mathbf{L}$, and $\mathbf{L}^H$ are the diagonal, strictly lower triangular, and strictly upper triangular components of $\mathbf{W}$, respectively.

\emph{Step 2:} Compute initial solution $\mathbf{s}^{(0)}$ of the GS method. Usually, $\mathbf{s}^{(0)}$ is set as a zero vector if no \emph{prior} information of the final solution is available.

\emph{Step 3:} The transmitted signal vector $\mathbf{s}$ is then estimated iteratively as follows:
\begin{equation}\label{eq:gs}
\mathbf{s}^{(k)}=(\mathbf{D+L})^{-1}(\mathbf{y}^\mathrm{MF}-\mathbf{L}^H\mathbf{s}^{(k-1)}),\; k=1,2,\ldots,K,
\end{equation}
where $K$ is the maximum number of iterations.
\subsection{Fast-Converging Initial Solution}
\label{subsec:ini}
As discussed in Section \ref{subsec:gs}, since no $\emph{a priori}$ information of the final solution is available, the initial solution $\mathbf{s}^{(0)}$ in \eqref{eq:gs} is often set as an all-zero vector. Such a choice is simple but requires a large number of iterations. In general, the initial solution plays an important role for the convergence of the GS method and affects both complexity and accuracy when the number of iterations is finite (and small).

The exact solution of GS method is given in Eq. \eqref{eq:detection}. Note that the inverse of the regularized Gram matrix $\mathbf{W}^{-1}$ here can also be computed by the following NSE:
\begin{equation}
\label{eq:neum}
\mathbf{W}^{-1}=\sum\nolimits_{k=0}^{\infty}(\mathbf{I}_{N_t}-\mathbf{X}^{-1}\mathbf{W})^k\mathbf{X}^{-1},
\end{equation}
where $\mathbf{X}^{-1}$ is an arbitrary matrix satisfying the condition
\begin{equation}
\lim_{k \rightarrow \infty}{(\mathbf{I}_{N_t}-\mathbf{X}^{-1}\mathbf{W})^k=\mathbf{0}_{N_t}}.
\end{equation}
Letting $\mathbf{X=D}$ for the sake of low complexity and keeping only the first $2$ terms of NSE, we arrive at the following $2$-term approximation of $\mathbf{W}^{-1}$ \cite{Wu2013iscas}
\begin{equation}
\label{eq:wtilde}
\mathbf{W}_2^{-1}=\mathbf{D}^{-1}-\mathbf{D}^{-1}\mathbf{E}\mathbf{D}^{-1},
\end{equation}
where $\mathbf{E}$ is the off diagonal part of $\mathbf{W}$.
Exploiting such an efficient approximation, we have the new initial solution $\mathbf{s}^{(0)}$ as follows:
\begin{equation}\label{eq:initsolu}
\mathbf{s}^{(0)}=\mathbf{W}_2^{-1}\mathbf{y}^\mathrm{MF}.
\end{equation}
\begin{Rem}
The NSE-based approximation is only efficient with the number of terms less than three, otherwise it will still suffer from high complexity.
\end{Rem}
\begin{algorithm}[tbp]
\caption{Improved GS detection for massive MIMO (IGS)}
\begin{algorithmic}[1]
\STATE \textbf{Input:} $\mathbf{y}$, $\mathbf{H}$, $N_0$ and $K$
\STATE (\textit{Preprocessing})
\STATE $\mathbf{W}=\mathbf{H}^H\mathbf{H}+N_0\mathbf{I}_{N_t}$ and $\mathbf{y}^\mathrm{MF}=\mathbf{H}^H\mathbf{y}$
\STATE $\mathbf{W}=\mathbf{D}+\mathbf{E}$ and $\mathbf{E}=\mathbf{L}+\mathbf{L}^H$
\STATE (\textit{Initialization})
\STATE $\mathbf{W}_2^{-1}=\mathbf{D}^{-1}-\mathbf{D}^{-1}\mathbf{E}\mathbf{D}^{-1}$
\STATE $\mathbf{s}^{(0)}=\mathbf{W}_2^{-1}\mathbf{y}^\mathrm{MF}$
\STATE (\textit{GS iteration})
\FOR{$k=1,2,...,K$}
\STATE $\mathbf{s}^{(k)}=(\mathbf{D+L})^{-1}(\mathbf{y}^\mathrm{MF}-\mathbf{L}^H\mathbf{s}^{(k-1)})$
\ENDFOR
\STATE (\textit{LLR computation})
\STATE $\mathbf{\widetilde{U}}=\mathbf{I}_{N_t}-N_0\mathbf{W}_2^{-1}$
\FOR{$i=1,2,...,N_t$}
\FOR{$b=1,2,...,B$}
\STATE $\hat{L}_{ib}=\rho_i\lambda_b(z_i)$ (See Eq. \eqref{eq:apprLLR} for details.)
\ENDFOR
\ENDFOR
\STATE \textbf{Output:}  $\hat{L}_{ib}$
\end{algorithmic}
\label{alg:igs}
\end{algorithm}
\subsection{Efficient LLR Computation}
\label{subsec:effillr}
Although the computational complexity of LLR computation has been reduced significantly in \eqref{eq:apprLLR}, it requires the inverse matrix $\mathbf{W}^{-1}$ to compute the effective channel gains $\mu_i$. For the purpose of computing LLRs more efficiently, we propose an approximated method to obtain the effective channel gains with a negligible performance loss. Firstly, the equivalent channel matrix can be rewritten as
\begin{equation}
\mathbf{U}=\mathbf{W}^{-1}\mathbf{G}=\mathbf{I}_{N_t}-N_0\mathbf{W}^{-1}.
\end{equation}
Inspired by the proposed initial solution in Section \ref{subsec:ini}, we also exploit the $2$-term NSE $\mathbf{W}_2^{-1}$ to approximate $\mathbf{W}^{-1}$ here:
\begin{equation}
\label{eq:utilde}
\mathbf{\widetilde{U}}=\mathbf{I}_{N_t}-N_0\mathbf{W}_2^{-1}.
\end{equation}
Therefore, the effective channel gain can be computed by
\begin{equation}
\label{eq:mui}
\tilde{\mu_i}=1-N_0W'_{ii},
\end{equation}
where $W'_{ii}$ denotes the $i$th diagonal entry of $\mathbf{W}_2^{-1}$.
Then, the approximated LLRs can be efficiently computed by \eqref{eq:apprLLR}.

\subsection{Computational Complexity Analysis}
Since most of the linear MMSE algorithms of massive MIMO detection require
to pre-compute the regularized Gram matrix $\mathbf{W}$ and the matched-filter output $\mathbf{y}^\mathrm{MF}$, we focus mainly on the complexity of the other parts. Therefore, we have: 
\begin{Lem}\label{lem:cpl}
The computational complexity of the proposed IGS algorithm scales as $\mathcal{O}((K+2)N_t^2)$.
\end{Lem}
\begin{figure}[t]
\centering
    \includegraphics[width=3 in]{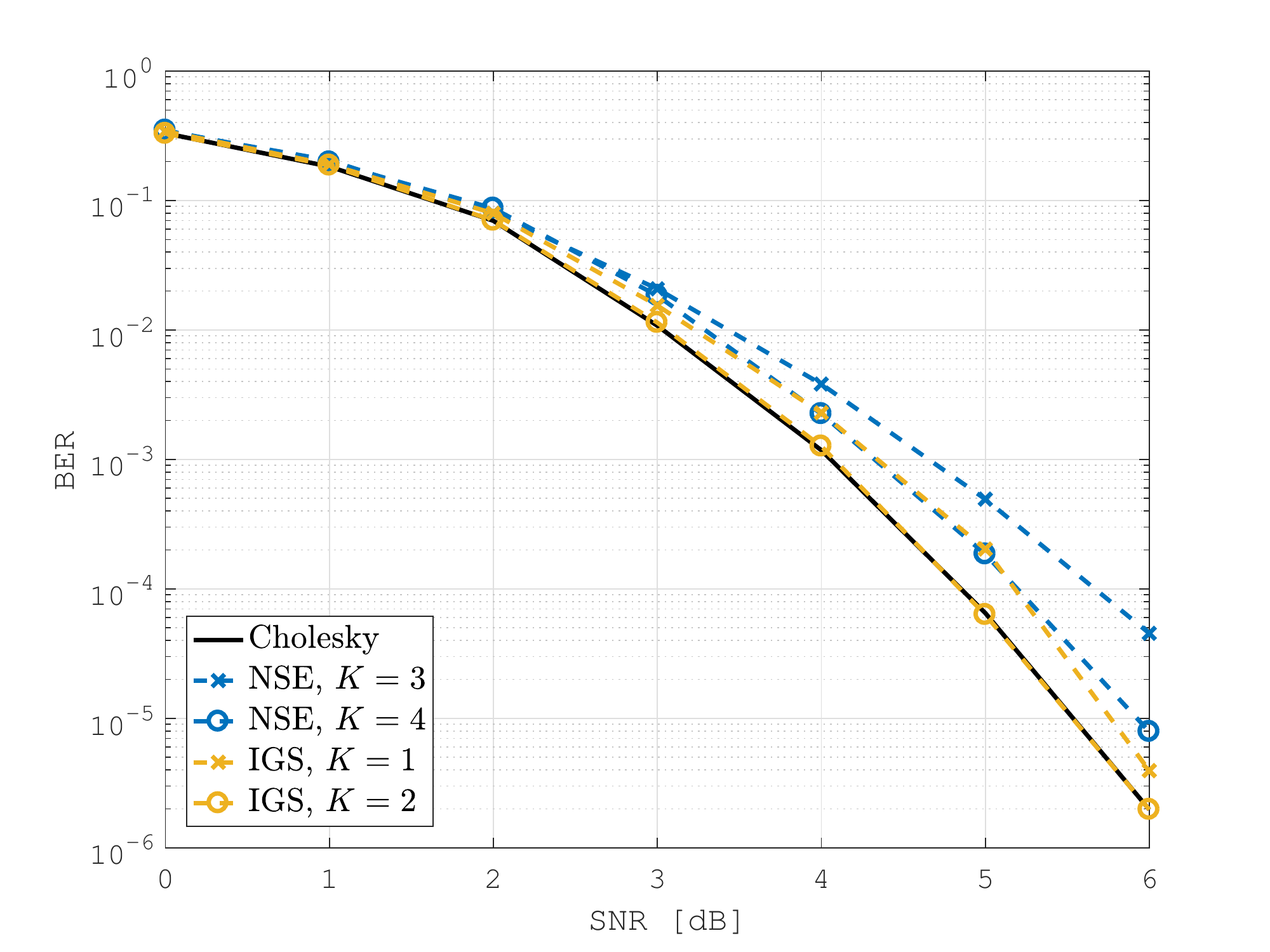}
    \caption{BER Performance comparison for different methods with $N_r=128$ and $N_t=16$.}
    \label{fig:ber_128} 
  \end{figure}

\begin{figure}[t]
\centering
\includegraphics[width=3 in]{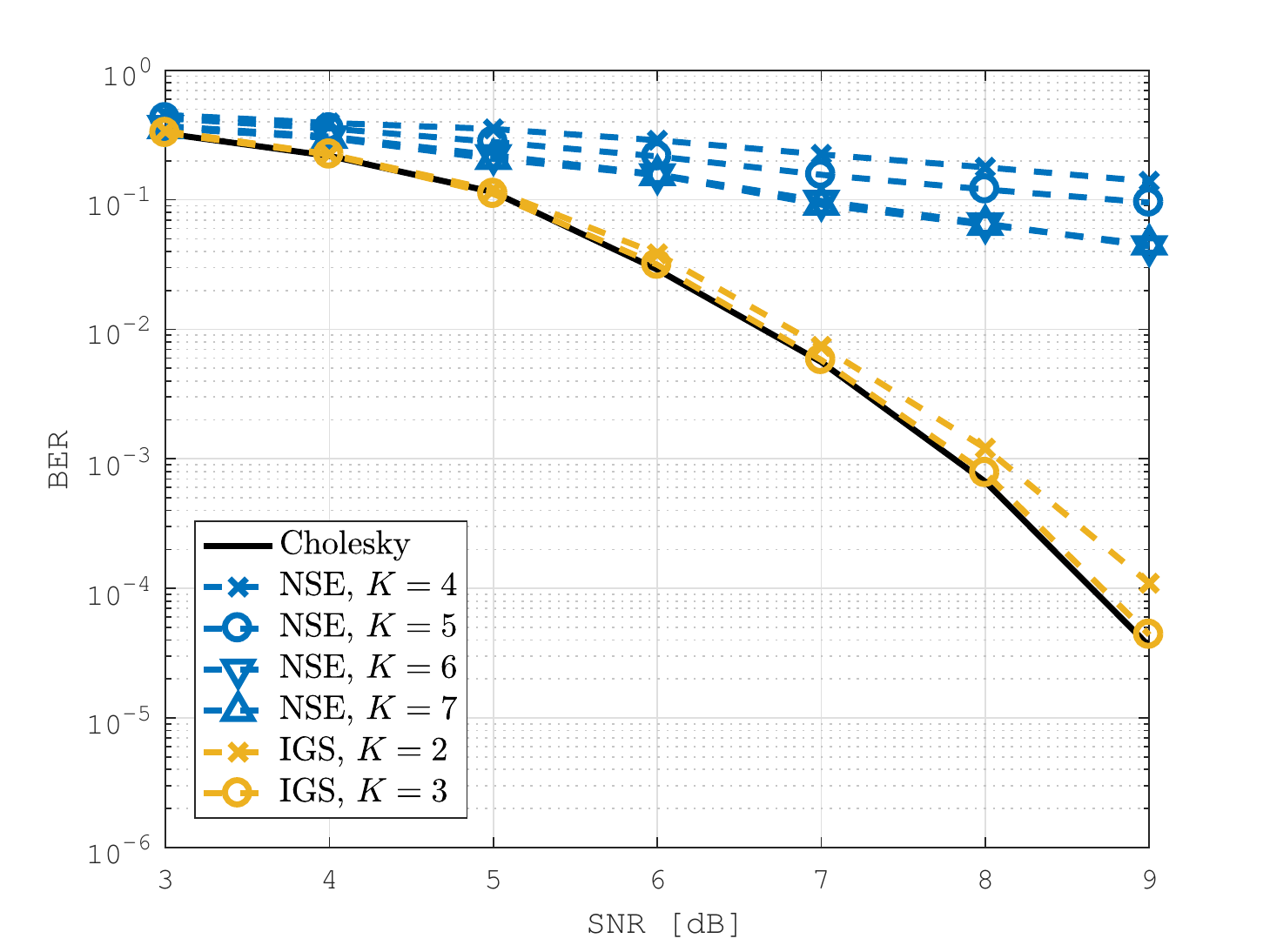}
  \caption{BER Performance comparison for different methods with $N_r=64$ and $N_t=16$.}
    \label{fig:ber_64} 
    \end{figure}

\begin{figure}[t]
\centering
\includegraphics[width=3 in]{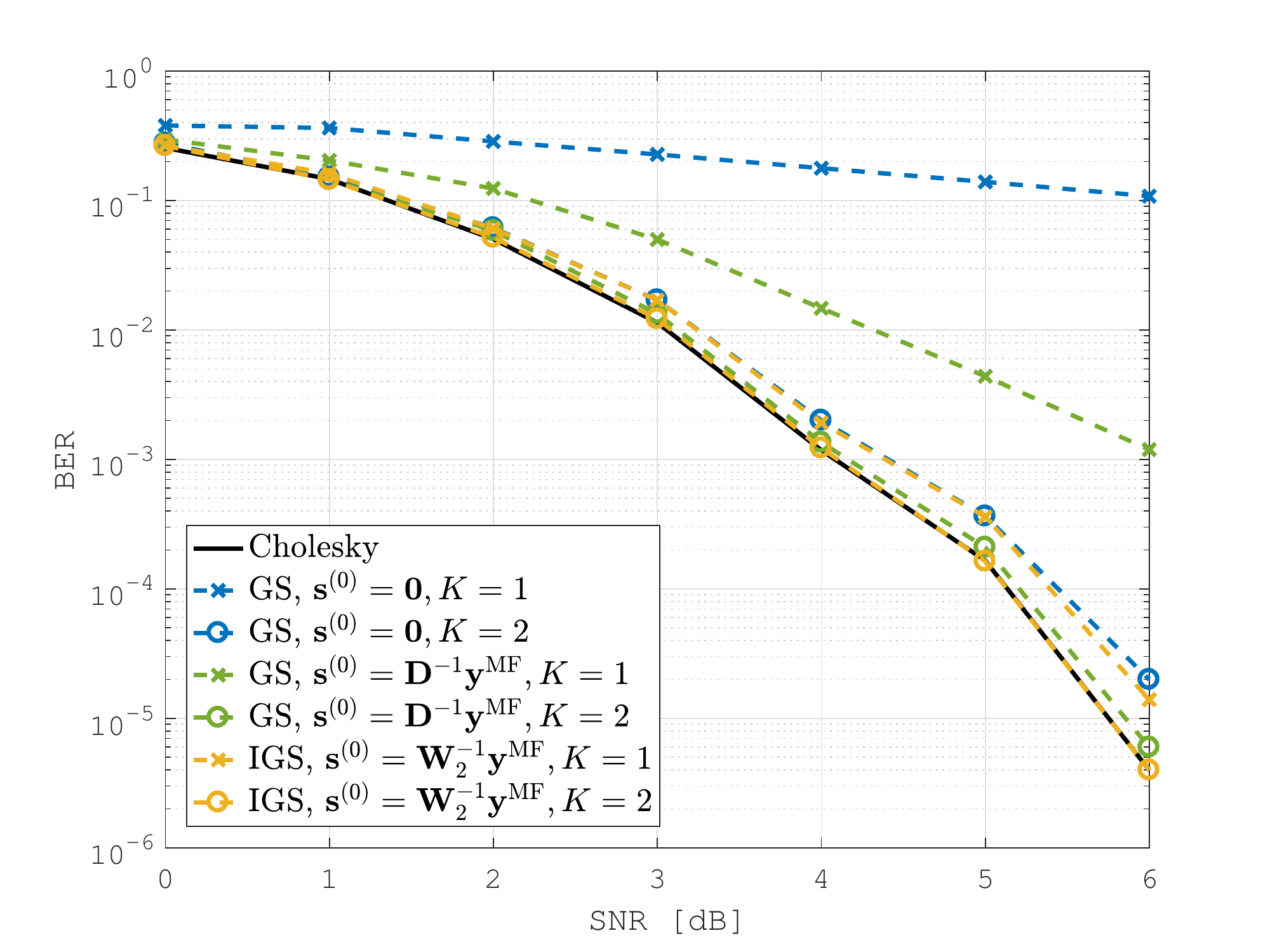}
\caption{BER performance comparison for different initial solutions with $N_r=64$ and $N_t=8$.}
\label{fig:difinit}
\end{figure}

\begin{figure}[t]
\centering
\includegraphics[width=3 in]{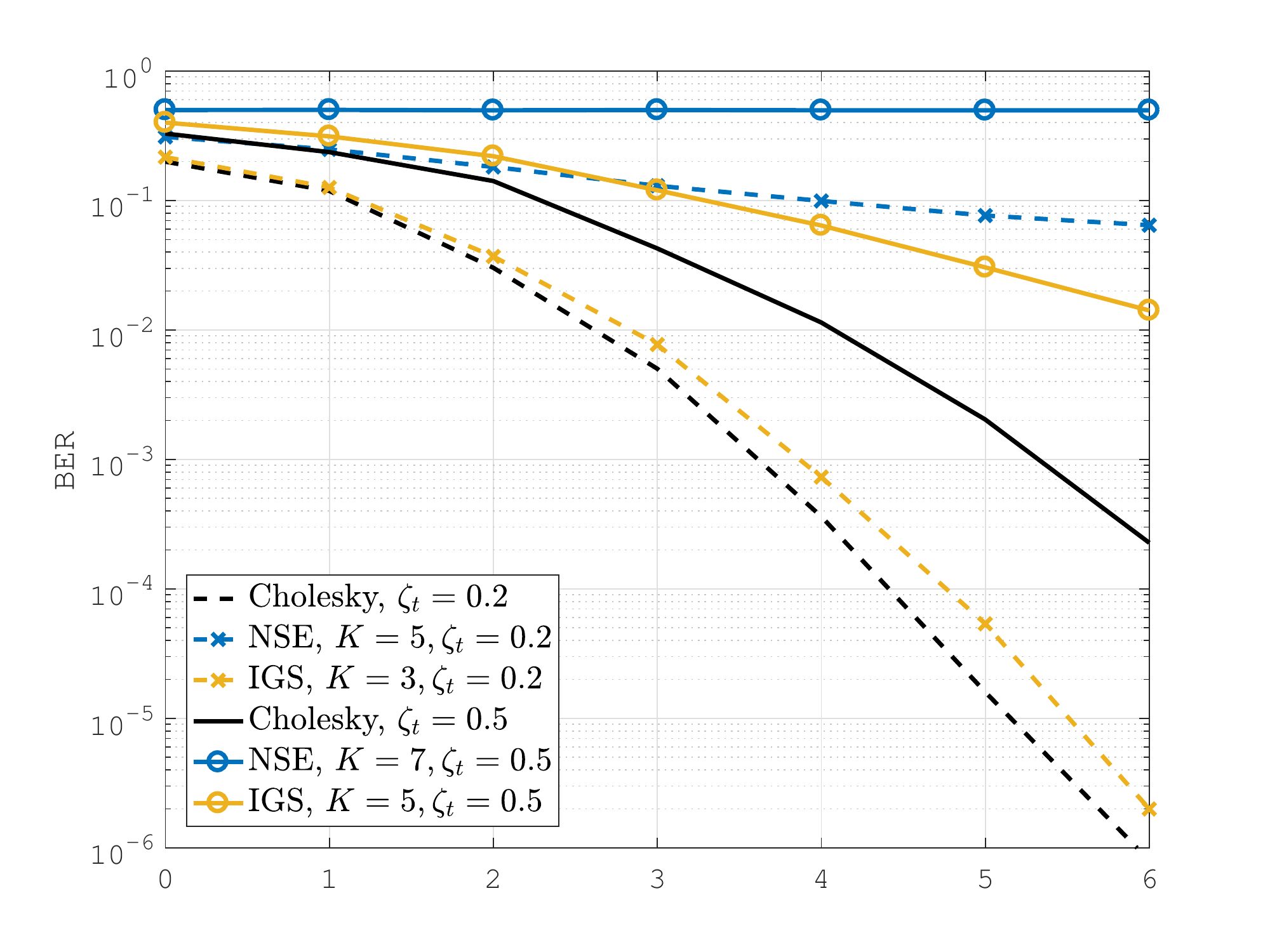}
\caption{BER performance comparison for different correlation factors $\zeta_t$ with
$N_r = 128$, $N_t = 16$ and $\zeta_r$ = 0.4.}
\label{fig:difcorr}
\end{figure}
\begin{IEEEproof}
In what follows, we estimate the computation complexity of reciprocal operations in term of complex-valued multiplications.

\emph{Computing $\mathbf{W}_2^{-1}$:}
As the multiplication of a $N_t \times N_t$ diagonal matrix and a $N_t \times N_t$ matrix requires $N_t^2$ complex-valued multiplications, the number of complex-valued multiplications for calculating $\mathbf{W}_2^{-1}$ is ($2N_t^2-N_t$) according to  \eqref{eq:wtilde}. Since
$\mathbf{D}^{-1}\mathbf{E}\mathbf{D}^{-1}$ is Hermitian, only the lower-triangle component or the upper-triangle component needs to be
computed. Hence, the total number of complex-valued multiplications is $N_t^2$.

\emph{Computing $\mathbf{s}^{(0)}$:}
As the matched-filter output $\mathbf{y}^\mathrm{MF}$ is a $N_t \times 1$ vector, it requires $N_t^2$ complex-valued multiplications to obtain the proposed initial solution $\mathbf{s}^{(0)}$ according to \eqref{eq:initsolu}.

\emph{Solving \eqref{eq:gs}:}
Considering $\mathbf{(D+L)}$ is a lower-triangular matrix, the computation of \eqref{eq:gs} after $K$ iterations requires $KN_t^2$ complex-valued multiplications. Note that since the proposed initial solution is relatively close to the exact solution in favourable propagation environments, $K$ could be quite small.

\emph{Computing $\mu_i$:}
Since $\mathbf{W}_2^{-1}$ has been computed, the required number of complex-valued multiplications of computing $\mu_i$ (for $i=1,\ldots,N_t$) is simply $N_t$ according to \eqref{eq:mui}.

To sum up, the overall required number of complex-valued multiplications of proposed IGS algorithm is $(K+2)N_t^2$. For massive MIMO detection, the computational complexity of IGS reduces to $\mathcal{O}((K+2)N_t^2)$.
\end{IEEEproof}
\subsection{Simulation Results}\label{subs:simu}
Numerical results of BER performance against SNR are shown in Figs. \ref{fig:ber_64}--\ref{fig:difcorr} to compare the proposed IGS algorithm with the \emph{Cholesky} decomposition, the NSE-based method, and conventional GS-based algorithms. We consider massive MIMO systems with different parameters , where a standard rate-$1/2$ convolutional channel code and $64$ quadrature amplitude modulation (QAM) scheme are employed. All the compared methods compute soft-outputs. At the receiver, LLRs are extracted from the detected signal for soft-input \emph{Viterbi} decoding.
\begin{figure*}[tbp]
\centering
\includegraphics[width=6in]{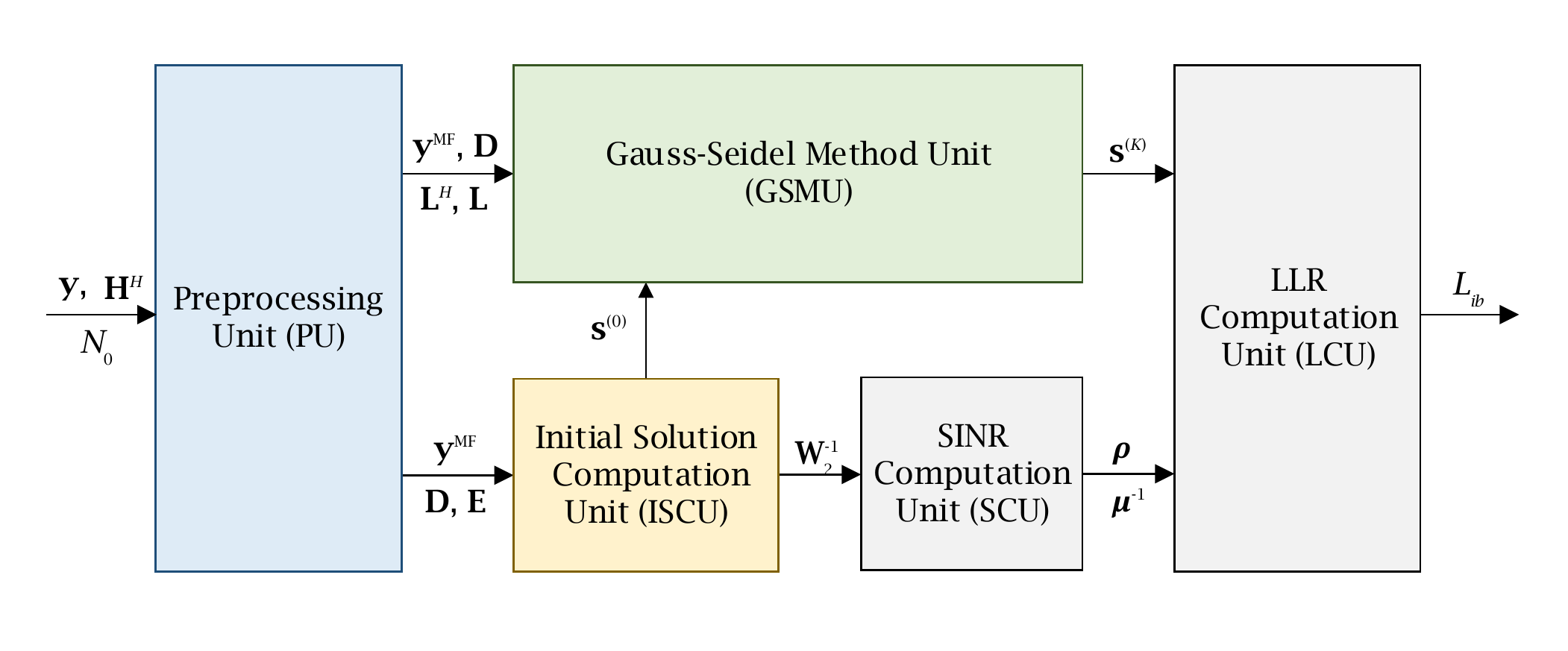}
\caption{High-level architecture of IGS detector.}
\label{fig:overview}
\end{figure*}
\begin{Rem}\label{rem:simu}
We let the number of iterations of the NSE-based approach greater than that of IGS ($K_\mathrm{NSE}=K_\mathrm{GS}+2$) in order to enable a fair comparison, since the 2-term NSE is employed to obtain the initial solution of IGS.
\end{Rem}
In Fig. \ref{fig:ber_128} and \ref{fig:ber_64}, we compare the BER performance of IGS with the other conventional
approaches under different antenna configurations. It is shown that IGS outperforms the NSE-based algorithm under both antenna configurations of $64\times 16$ and $128\times 16$. According to Fig. \ref{fig:ber_128}, IGS with $K=1$ achieves similar performance as the NSE-based algorithm with $K=4$. Noting that with the greater system loading factor $N_t/N_r$ (shown in Fig. \ref{fig:ber_64}), the NSE-based algorithm converges much slower than IGS. Therefore, the proposed IGS algorithm has advantage of convergence rate over the NSE-based one.

In Fig. \ref{fig:difinit}, we compare the performance of the GS-based algorithms with different initial solutions. Compared to existing choices, the proposed initial solution successfully accelerates the convergence. It is shown that the GS method with the proposed initial solution for $K=1$ almost achieves the same performance of that with the zero-vector initial for $K=2$. Moreover, IGS also outperforms the one with initial solution $\mathbf{s}^{(0)}=\mathbf{D}^{-1}\mathbf{y}^\mathrm{MF}$ proposed in \cite{Dai2015} in terms of error-rate performance for given $K$.

In Fig. \ref{fig:difcorr}, we study the BER performance of IGS and other conventional approaches considering the spatial correlation of realistic MIMO systems. Here we adopt the \emph{Kronecker} model proposed in \cite{Godana2013}, where $\zeta_r$ and $\zeta_t$ $(0 \leq \zeta \leq 1)$ denote the correlated factor at the BS and the user sides respectively. We can see that all these approaches degrade to various extents as the channel correlation becomes serious. For both $\zeta_t=0.2$ and $\zeta_t=0.5$, the conventional NSE-based algorithm is hardly able to converge. However, IGS still converges with relatively small number of iterations.

\section{VLSI Architecture}\label{sec:arch}
In this section, we describe a low-complexity VLSI architecture for the proposed IGS algorithm and provide several solutions to reduce hardware overhead and processing latency.
\begin{figure}[tbp]
\includegraphics{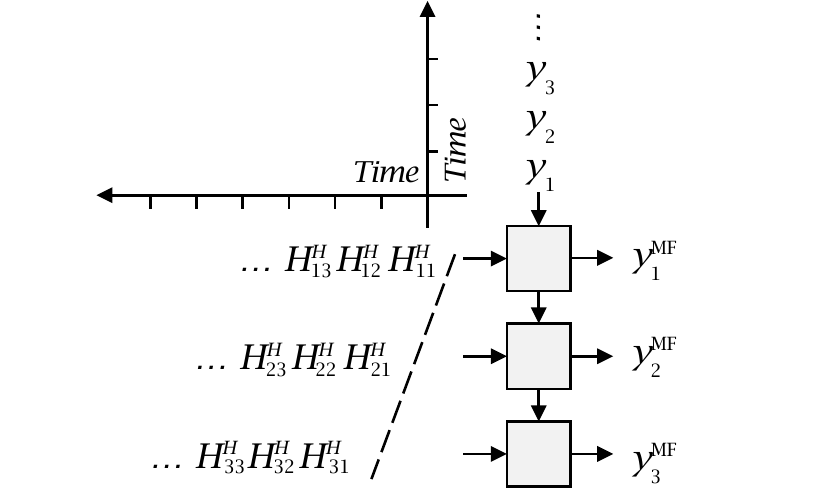}
\centering
\caption{Array structure of matched filter (e.g., $N_t=3$).}
\label{fig:mf}
\end{figure}
\begin{figure}[tbp]
\centering
\includegraphics{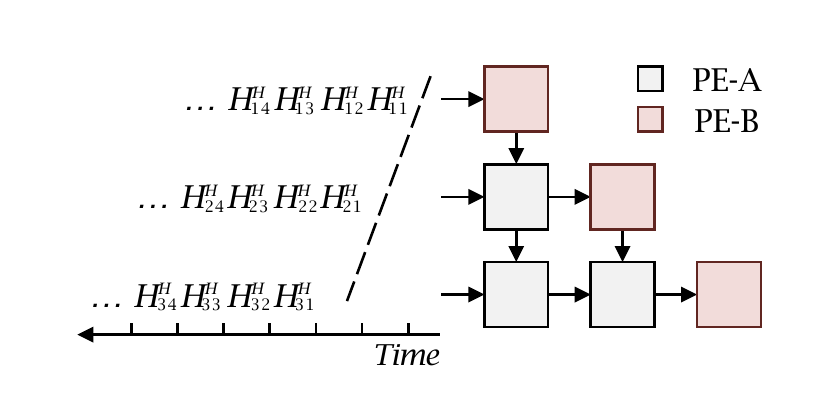}
\caption{Array structure of regularized Gram matrix computation unit (e.g., $N_t=3$).}
\label{fig:gram}
\end{figure}
\subsection{Architecture Overview}
The VLSI architecture is depicted in Fig. \ref{fig:overview}. The proposed architecture consists of five units:
\begin{inparaenum}
\item{preprocessing unit (PU)},
\item{initial solution computation unit (ISCU)},
\item{GS method unit (GSMU)},
\item{SINR computation unit (SCU)}, and
\item{LLR computation unit (LCU)}.
\end{inparaenum}
Fed by $\mathbf{y}$, $\mathbf{H}^H$ and $N_0$,
PU performs matched filtering $\mathbf{y}^\mathrm{MF}=\mathbf{H}^H\mathbf{y}$ and computes the regularized Gram matrix $\mathbf{W}$. Note that both operations can be performed by systolic arrays to achieve high-throughput. The outputs of PU are then passed to ISCU for computing the 2-term NSE $\mathbf{W}^{-1}_2$ and the initial solution of GS method $\mathbf{s}^{(0)}$. While GSMU is iteratively computing $\mathbf{\hat{s}}$, SCU is computing the equivalent channel gain $\mu_i$ and the post-equalization SINR $\rho_i$. Finally, LCU performs the computation of LLRs according to Eq. \eqref{eq:apprLLR}.
\subsection{Preprocessing Unit (PU)}
The preprocessing unit is employed to compute the matched filter output $\mathbf{y}^\mathrm{MF}$ and the regularized Gram matrix $\mathbf{W}$, which will be further passed to other units. Note that this unit is able to perform the operations above in parallel, since there is no data dependence between them.
\begin{figure}[tbp]
\centering
\includegraphics[width=3in]{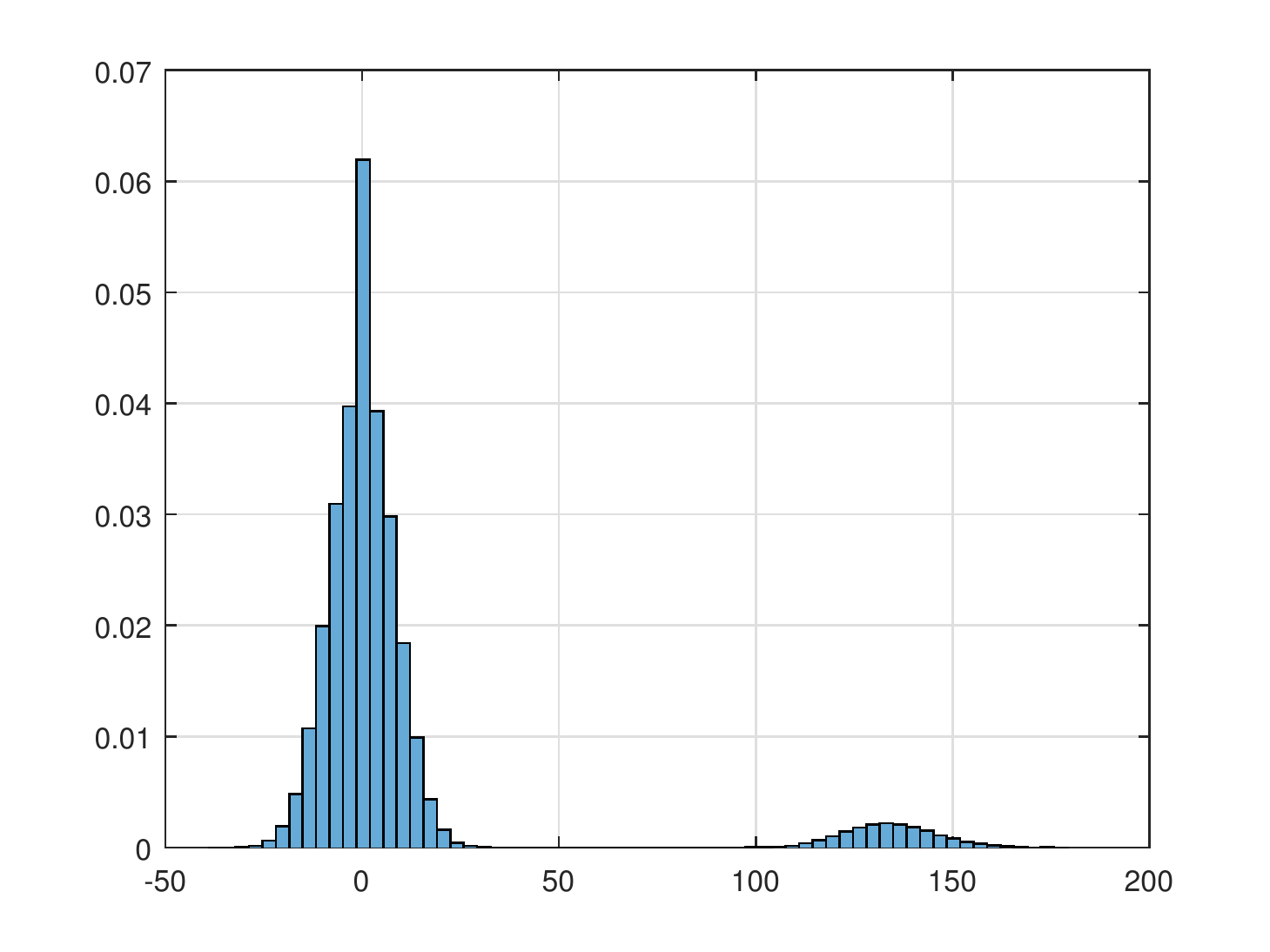}
\caption{Distribution of the value in regularized Gram matrix $\mathbf{W}$ with $N_r=128$ and $N_t=8$.}
\label{fig:w_dist}
\end{figure}
\subsubsection{Matched Filtering}
The matched filter (MF) consists of a linear array of $N_t$ PEs and performs the operation of $\mathbf{y}^\mathrm{MF}=\mathbf{H}^H\mathbf{y}$. Fig. \ref{fig:mf} shows the systolic array structure of MF. In each clock cycle, MF reads a new entry of $\mathbf{y}$ and the corresponding entries of $\mathbf{H}^H$, and the multiply-accumulate (MAC) operation is performed in each PE.
\begin{Rem}
The total processing latency of MF is ($N_t+N_r-1$), and it utilizes $N_t$ complex-valued multipliers.
\end{Rem}
\begin{figure}[bp]
\centering
\includegraphics{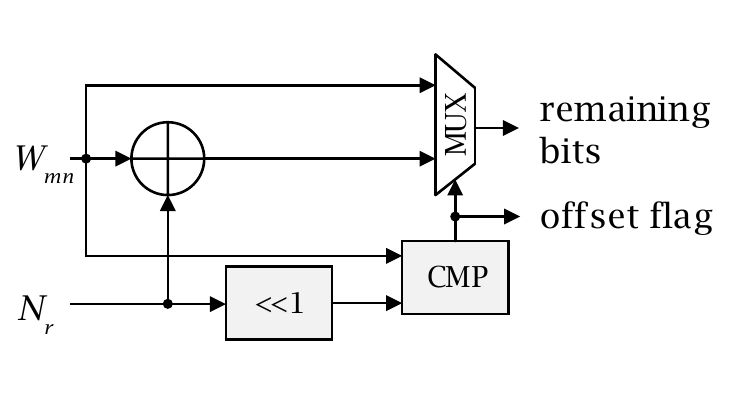}
\caption{Architecture for proposed hardware-efficient quantizer.}
\label{fig:comp_arch}
\end{figure}
\subsubsection{Regularized Gram Matrix Computation}
In massive MIMO systems, the dimension of the Gram matrix $\mathbf{G}$ tends to be very large. Therefore, the conventional systolic array for matrix-matrix multiplication which consists of $N_t \times N_t$ PEs is not scalable for large $N_t$. As discussed in Section \ref{subsec:gs}, the Gram matrix in massive MIMO uplink is Hermitian positive definite. Hence, either upper triangular part or lower triangular of $\mathbf{G}$ is required to be calculated. Fig. \ref{fig:gram} depicts the systolic array structure for computing the regularized Gram matrix (RGM).
In this array, only $N_t(N_t+1)/2$ PEs are employed to compute the lower triangular part of the Gram matrix $\mathbf{G}$. Each PE in the array contains a MAC unit.
There are two types of PEs in the array, denoted by PE-A and PE-B, respectively.
The transposed channel matrix $\mathbf{H}^H$ is shifted one column at a time into the systolic array. PE-As have the same structure as PEs of MF. Once an input value reaches a PE-B, the value is
conjugated and passed to the lower part of the systolic array. Then, PE-Bs add the noise variance $N_0$ to the diagonal entries of $\mathbf{G}$. Finally, the lower triangular part of the regularized Gram matrix $\mathbf{L}$, the upper triangular part $\mathbf{L}^H$ and the diagonal part $\mathbf{D}$ are stored in the register files.
\begin{Rem}
Total processing latency of RGM is ($2N_t+N_r-1$), and it uses $N_t(N_t-1)/2$ complex-valued multipliers and $N_t$ real-valued multipliers since the diagonal entries of $\mathbf{W}$ are real-valued.
\end{Rem}
\begin{figure}[tbp]
\centering
\includegraphics{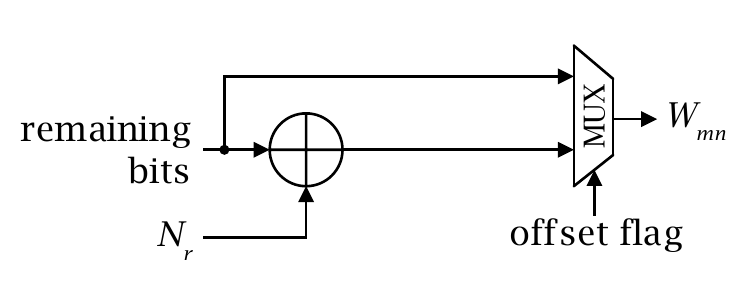}
\caption{Architecture for data decompression.}
\label{fig:decomp_arch}
\end{figure}
\begin{figure}[bp]
\centering
\includegraphics{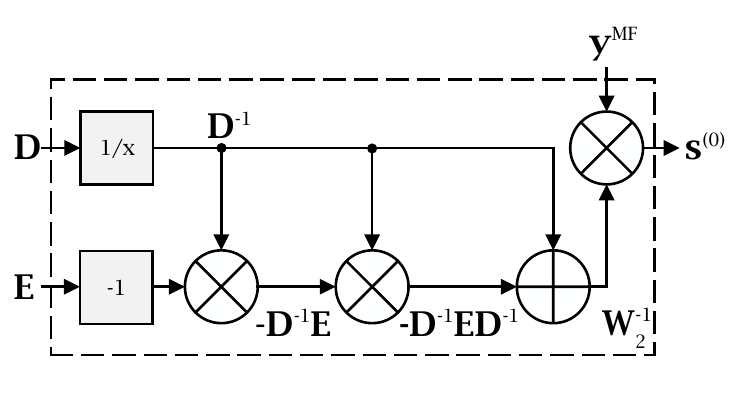}
\caption{Data flow graph of initial solution computation.}
\label{fig:nse_arch_hl}
\end{figure}
\subsubsection{Data Compression Scheme}
\label{subsubsec:quan}
Note that the regularized Gram matrix $\mathbf{W}$ is diagonally dominant in the case of massive MIMO with i.i.d. assumption, meaning that the diagonal entries of $\mathbf{W}$ are much greater than the off-diagonal ones. Hence, conventional uniform quantization schemes that cover the entire dynamic range of entries of $\mathbf{W}$ will cost a mass of hardware resources. Fig. \ref{fig:w_dist} shows the distribution of entries of $\mathbf{W}$ in form of histogram. Obviously, values of $\mathbf{W}$ can be separated into two groups, one is around zero and the other is around $N_t$. By exploiting this property, the hardware overhead for storing and processing $\mathbf{W}$ can be saved significantly.
The hardware-efficient quantization scheme with data compression for $\mathbf{W}$ can be denoted as follows (see Fig. \ref{fig:comp_arch} for architecture details)

\emph{Step 1:} Compare $W_{mn}$ with $N_t/2$. If the former is bigger than the latter, set the first bit of the fixed-point output (offset flag) as $1$, otherwise set it as $0$.

\emph{Step 2:} If $W_{mn} > N_t/2$, subtract $W_{mn}$ with $N_t$ and then obtain the remaining bits of fixed-point output.

The compressed bits of $W_{mn}$ will be sent to other units in detector, e.g. ISCU and GSMU. Before these units start to compute, the corresponding operation of data decompression is required. As shown in Fig. \ref{fig:decomp_arch}, the procedure of data decompression is simple and straightforward.

\subsection{Initial Solution Computation Unit (ISCU)}
As shown in Fig. \ref{fig:nse_arch_hl}, ISCU first performs the approximate matrix inversion $\mathbf{W}_2^{-1}$ and then computes the initial solution $\mathbf{s}^{(0)}$ of GS method. Since the computation of $\mathbf{s}^{(0)}$ is a typical matrix-vector multiplication, which can be performed by an array of MACs similar to MF, we focus mainly on the architecture design of the 2-term NSE approximation $\mathbf{D}^{-1}-\mathbf{D}^{-1}\mathbf{E}\mathbf{D}^{-1}$.
\begin{figure}[tbp]
\centering
  \subfigure[Basic architecture for 2-term NSE.]{
    \label{fig:nse_a} 
    \includegraphics{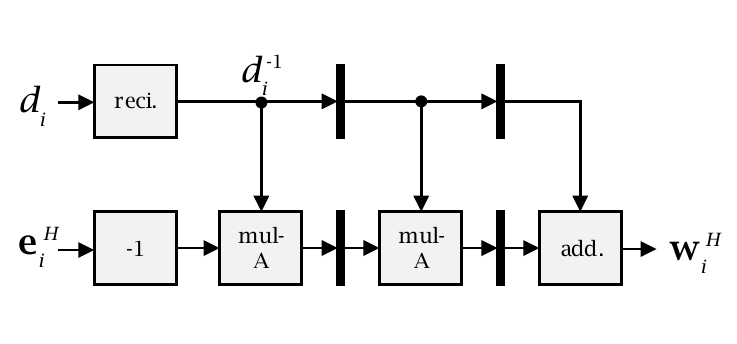}}
  \subfigure[Proposed architecture for 2-term NSE with low latency.]{
    \label{fig:nse_b} 
    \includegraphics{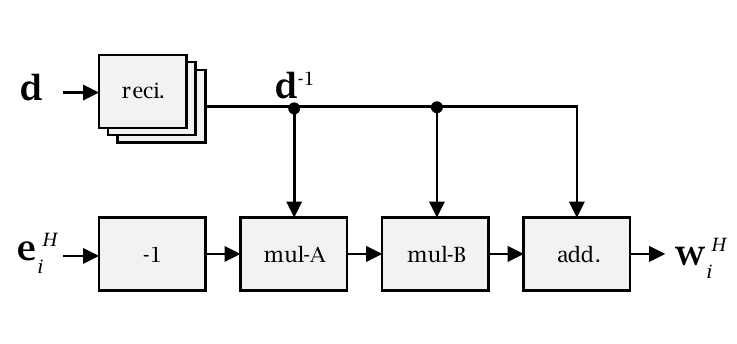}}
  \caption{Two architectures for 2-term NSE computation.}
  \label{fig:NSEarch} 
\end{figure}
A straightforward method to obtain $\mathbf{W}_2^{-1}$ is shown in Fig. \ref{fig:nse_a}, where $d_i$ is the $i$th diagonal entry of $\mathbf{D}$, $\mathbf{e}^H_i$ and $\mathbf{w}^H_i$ denote the $i$th row of $\mathbf{E}$ and $\mathbf{W}_2^{-1}$, respectively. To compute $\mathbf{D}^{-1}$, a reciprocal module based on look-up table (LUT), which is suitable for FPGA implementation, is employed. The diagonal matrix $\mathbf{D}^{-1}$ is stored as an vector $\mathbf{d}^{-1}$ for the sake of saving storage space. The function of mul-A is to multiply a vector by a scalar.
\begin{Rem}
For the architecture in Fig. \ref{fig:nse_a}, either one row or column is computed in mul-A per clock cycle, therefore four buffers are used to obtain a row of $\mathbf{W}_2^{-1}$ per clock cycle.
\end{Rem}

The major drawback of the architecture in Fig. \ref{fig:nse_a} is the high processing latency. Considering the regularized Gram matrix $\mathbf{W}$ of order $N_t$, an undesirable latency of $2N_t$ clock cycles are required to obtain $\mathbf{w}^H_i$. Since $\mathbf{W}=\mathbf{D}+\mathbf{E}$ and $\mathbf{W}$ is Hermitian, the off-diagonal matrix $\mathbf{E}$ is therefore Hermitian. Then, a low-latency version of this architecture is provided in Fig. \ref{fig:nse_b}. Instead of computing the reciprocal of $d_i$ per clock, the proposed structure in Fig. \ref{fig:nse_b} computes all entries of $\mathbf{D}$ during one clock cycle. Note that the mul-B is employed here to perform element-wise multiplication of $\mathbf{m}_i^H$ and $\mathbf{d}^{-1}$.
\begin{Rem}
By exploiting the Hermitian characteristic, the proposed structure in Fig. \ref{fig:nse_b} is able to obtain $\mathbf{W}$ in $N_t$ clock cycles, and extra buffers are no longer needed.
\end{Rem}

After obtaining the 2-term NSE $\mathbf{W}_2^{-1}$, the initial solution of the GS method is computed according to Eq. \ref{eq:initsolu}, which can be performed by the systolic array in Fig. \ref{fig:mf}. It requires ($2N_t-1$) clock cycles to perform this operation.
\begin{figure}[tbp]
\centering
  \subfigure[Basic architecture of GS iteration.]{
    \label{fig:gs_a} 
    \includegraphics{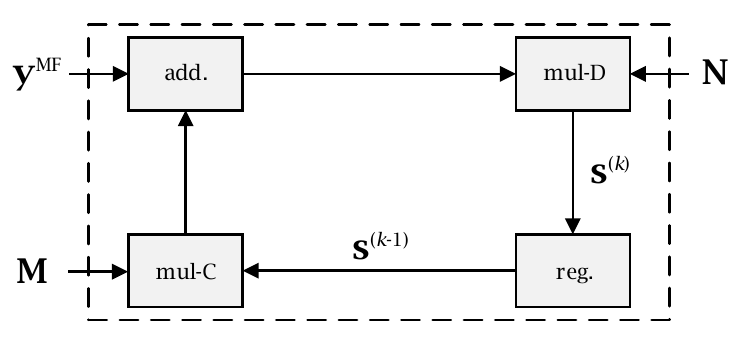}}
  \subfigure[Hardware-efficient architecture of GS iteration.]{
    \label{fig:gs_b} 
    \includegraphics{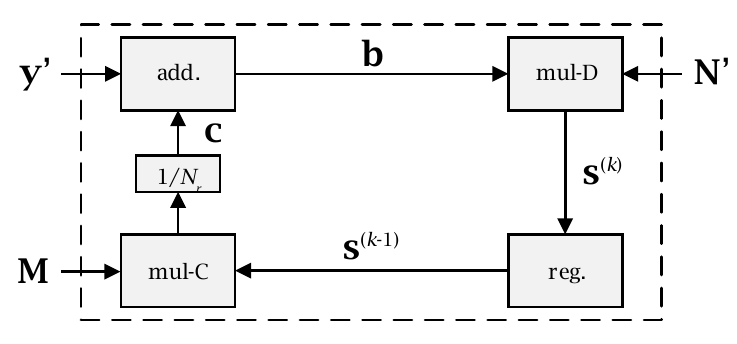}}
  \caption{Two architectures for GS iterative method.}
  \label{fig:GSarch} 
\end{figure}
\subsection{GS Method Unit (GSMU)}
Fig. \ref{fig:gs_a} shows the basic architecture of GS method unit, where $\mathbf{M}=-\mathbf{L}^H$ and $\mathbf{N}=(\mathbf{D}+\mathbf{L})^{-1}$. Since $(\mathbf{D}+\mathbf{L})$ is a lower triangular matrix, a specific systolic array that performs forward substitution (FS) \cite{Gaston1990} is employed here to compute $\mathbf{N}'$. According to Eq. (\ref{eq:gs}), each iteration of GS method can be divided into three phases. In the first phase, a systolic array for matrix-vector multiplication (denoted as mul-C) is employed to perform $-\mathbf{L}^H\mathbf{s}^{(k-1)}$.
In the second phase, the operation $\mathbf{b}=\mathbf{y}^\mathrm{MF}-\mathbf{L}^H\mathbf{s}^{(k-1)}$ is performed by $N_t$ complex-valued adders.
In the third phase, another systolic array denoted as mul-D is used to compute the matrix-vector multiplication of $\mathbf{N}'$ and $\mathbf{b}$.
These phases can be repeated for a configurable number of iteration until the error rate performance meets the requirement. Since $\mathbf{s}$ is updated from the previous iteration, only a few registers are required to store the latest $\mathbf{s}^{(k-1)}$.
\begin{Rem}
Both mul-C and mul-D in Fig. \ref{fig:GSarch} consist of $N_t$ MACs.
\end{Rem}
\subsubsection{Hardware-Efficient Architecture}
Fig. \ref{fig:gs_b} depicts the proposed hardware-efficient architecture of GS method, derived from the basic architecture in Fig. \ref{fig:gs_a}. In order to reduce the dynamic range of the values in GS method unit, we firstly normalize the input $\mathbf{y}^\mathrm{MF}$ and $\mathbf{N}$ into $\mathbf{y}'=N_r^{-1}\mathbf{y}^\mathrm{MF}$ and $\mathbf{N}'=N_r\mathbf{N}$. Such a trick is commonly used in fixed-point arithmetic. Therefore, $\mathbf{M}\mathbf{s}^{(k-1)}$ is scaled by $1/N_r$ correspondingly, ensuring that the final result is equivalent. Note that $N_r$ is usually a power of $2$, therefore these scaling operations can be performed easily by shifting. After scaling, word-length of the data in GSMU can be significantly shortened and therefore the hardware overhead is reduced. Since the inputs $\mathbf{L},\mathbf{D}$ and $\mathbf{L}^H$
are compressed according to Section \ref{subsubsec:quan}, they should be converted into conventional fixed-point numbers before computing. Also, the proposed architecture is irrelevant of the number of iterations, and therefore suitable for various applications.
\begin{figure}[tbp]
\centering
\includegraphics[width=3.4in]{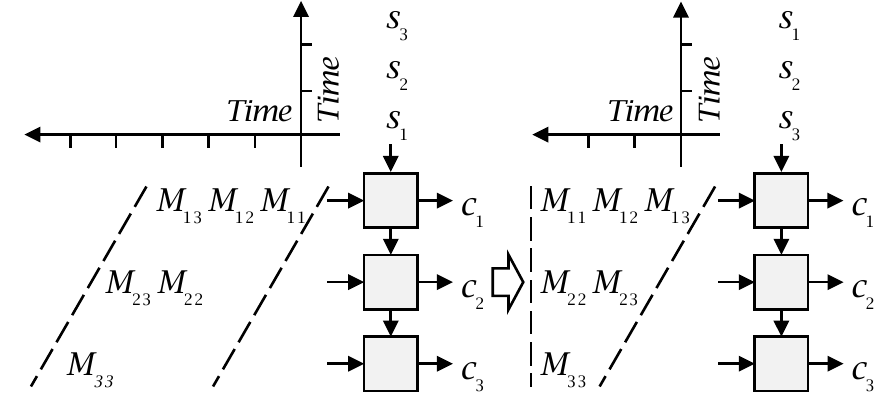}
\caption{Timing schedule of mul-C.}
\label{fig:gs_timing_1}
\end{figure}
\begin{figure}[tbp]
\centering
\includegraphics[width=3.4in]{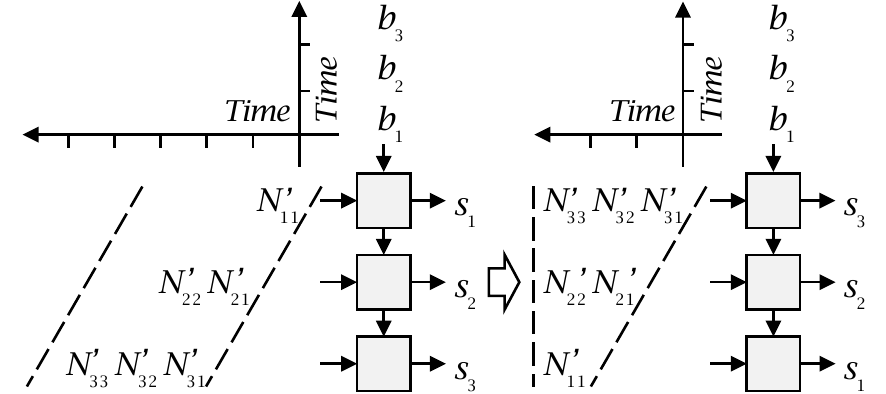}
\caption{Timing schedule of mul-D.}
\label{fig:gs_timing_2}
\end{figure}
\subsubsection{Timing Schedule for Lower Latency}
Usually, for a systolic array which performs matrix-vector multiplication of dimension $N_t$, it requires ($2N_t-1$) clock cycles to obtain the result (see the left parts of Fig. \ref{fig:gs_timing_1} and Fig. \ref{fig:gs_timing_2}).
It is worth noting that the GS iteration could have certain disadvantages because the variables that depend on each other can only be updated sequentially. Thus, it is important to reduce the overall latency of GS iterations so as to meet the throughput requirement. To this end, we propose a fast-converging initial solution in Section \ref{subsec:ini}, which is able to significantly reduce the number of iterations. Here, we introduce an efficient timing schedule scheme to further reduce the latency within each GS iteration.

As shown in Fig. \ref{fig:gs_timing_1}, since $\mathbf{M}=-\mathbf{L}^H$ is an upper-triangular matrix, we reverse the input sequence of $\mathbf{d}$ and each row of $\mathbf{M}$. As a result, the total latency of mul-C is reduced from ($2N_t-1$) clock cycles to only $N_t$ clock cycles. Likewise, since $\mathbf{N}'=N_r\mathbf{(D+L)}^{-1}$ is a lower-triangular matrix, we simply reverse the input sequence of $\mathbf{N}'$ (see Fig. \ref{fig:gs_timing_2}) and therefore reduce the total latency of mul-D into $N_t$ clock cycles.
\begin{Rem}
By exploiting the rescheduling schemes in Fig. \ref{fig:gs_timing_1} and \ref{fig:gs_timing_2}, latency of each GS iteration can be significantly reduced to half its original time.
\end{Rem}
\subsection{SCU and LCU}
The design of SCU is simple and efficient. According to Section \ref{subsec:effillr},
the equivalent channel matrix $\mathbf{U}$ can be easily approximated by adders and scalar multipliers. Then the SINR computation is simply carried out by multipliers and reciprocal modules as described in Section \ref{subsec:sommse}. Given the effective channel gains $\mu_i$ and the post-equalization SINR values $\rho_i$, the computation of the max-log LLRs can be simplified with Gray mappings according to Section \ref{subsec:sommse}. Hence, the LCU focuses mainly on evaluating the linear function $\lambda_b(z_i)$. The readers can refer to \cite{Sun2015} for more details of LCU architectures.

\subsection{Timing Schedule of IGS Detector}
\begin{figure}[tbp]
\centering
\includegraphics{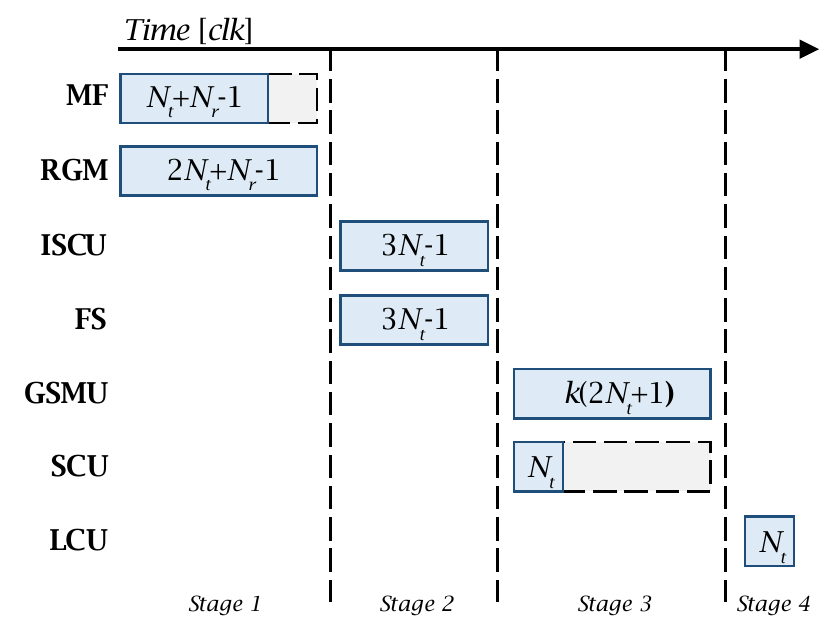}
\caption{Timing diagram for parallel implementation of IGS detector.}
\label{fig:sys_timing}
\end{figure}

As discussed before, because the GS method is inherently sequential, it is difficult to be performed in parallel. However, the proposed IGS detector consists of several units and some of them exist no data dependency. Therefore, some operations can be performed concurrently. After careful analysis and arrangement, the overall timing schedule of IGS detector is shown in Fig. \ref{fig:sys_timing}.

\begin{table*}[htbp]
\tabcolsep 2mm
\renewcommand{\arraystretch}{1.2}
\footnotesize
\centering
\caption{FPGA implementation results and comparison for $128 \times 8$ MIMO system.}
\label{tbl:results}
\begin{tabular}{llllll}
\Xhline{1.0pt}
\multirow{2}{*}{Detector} & \multirow{2}{*}{This work} & M. Wu \cite{Wu2014nse} & B. Yin \cite{Yin2015iscas} & O. Casta\~{n}eda \cite{Castaneda2016iscas} & M. Wu \cite{Wu2016cd}\\
& & [JSTSP'Nov. 14] & [ISCAS'15] & [ISCAS'16] & [TCAS-I'Dec. 16] \\ \hline
\rowcolor{mygray}
Method & IGS & NSE & CGLS & TASER & OCD \\
Performance & near-MMSE & near-MMSE & near-MMSE & near-ML & near-MMSE \\
\rowcolor{mygray}
Iteration \# & $1$ & $3$ & $3$ & $3$ & $3$\\
Modulation & $64$-QAM & $64$-QAM & $64$-QAM & QPSK & $64$-QAM\\
\rowcolor{mygray}
Instance or subcarrier \# & $10$ & $8$ & $1$ & N.A. & $24$\\ \hline
Slices & $35,721$ & $48,244$ & $1,094$ & $4,350$ & $11,094$\\
\rowcolor{mygray}
LUTs	& $105,135$    & $148,797$  & $3,324$ & $13,779$ & $23,914$\\
FFs   & $73,130$    &  $161,934$  & $3,878$ & $6,857$ & $43,008$ \\
\rowcolor{mygray}
DSP48s & $1,850$   & $1,016$  & $33$ & $168$ & $774$ \\
Block RAMs & $20$   & $16$  & $1$  & $0$ & $2$ \\ \hline
\rowcolor{mygray}
Latency [clks] & $202$ & $196$  & $951$ & $225$ & $795$\\
Maximum clock freq. [MHz] & $308$  & $317$  &$412$ & $72$ & $258$\\
\rowcolor{mygray}
Throughput [Mb/s] & $732$  & $621$ & $20$ & $50$ & $376$\\ \hline
Throughput/LUTs & $6,943$ & $4,173$ & $6,017$ & $3,629$ &$15,597$\\
\rowcolor{mygray}
Throughput/FFs & $9,982$  & $3,835$ & $5,157$ & $7,292$ & $8,743$\\
\Xhline{1.0pt}
\end{tabular}
\end{table*}


\section{Implementation Results}\label{sec:imple}
In this section, the proposed IGS detector has been implemented with Xilinx Virtex-7 XC7VX690T FPGA. The corresponding fixed-point parameters, FPGA implementation results, and comparison to other designs are provided here.
\begin{figure*}[tbp]
\centering
  \subfigure[$N_r=128$ and $N_t=8$.]{
    \label{fig:fp128x8} 
    \includegraphics[width=3 in]{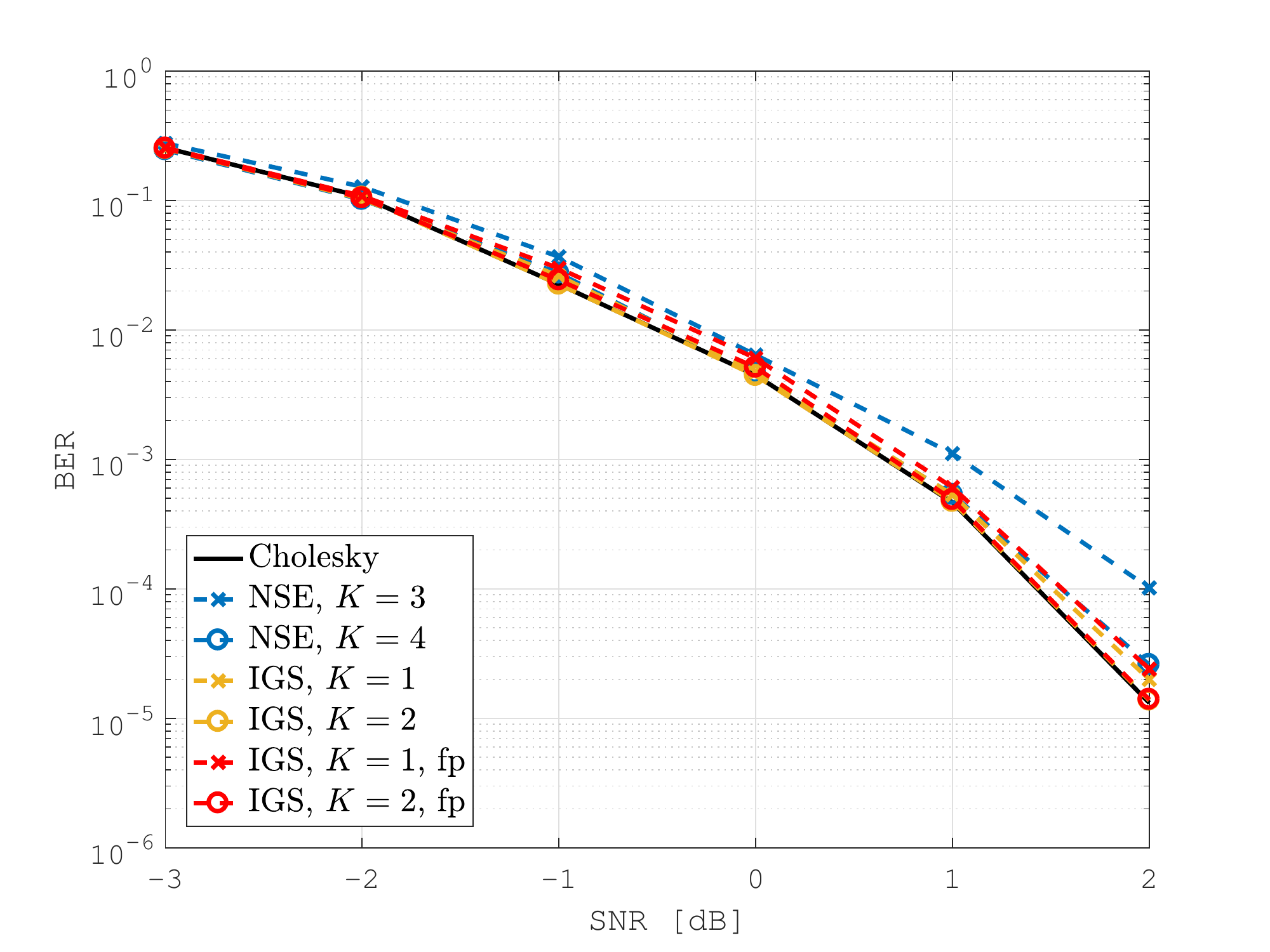}}
  \subfigure[$N_r=64$ and $N_t=8$.]{
    \label{fig:fp64x8} 
    \includegraphics[width=3 in]{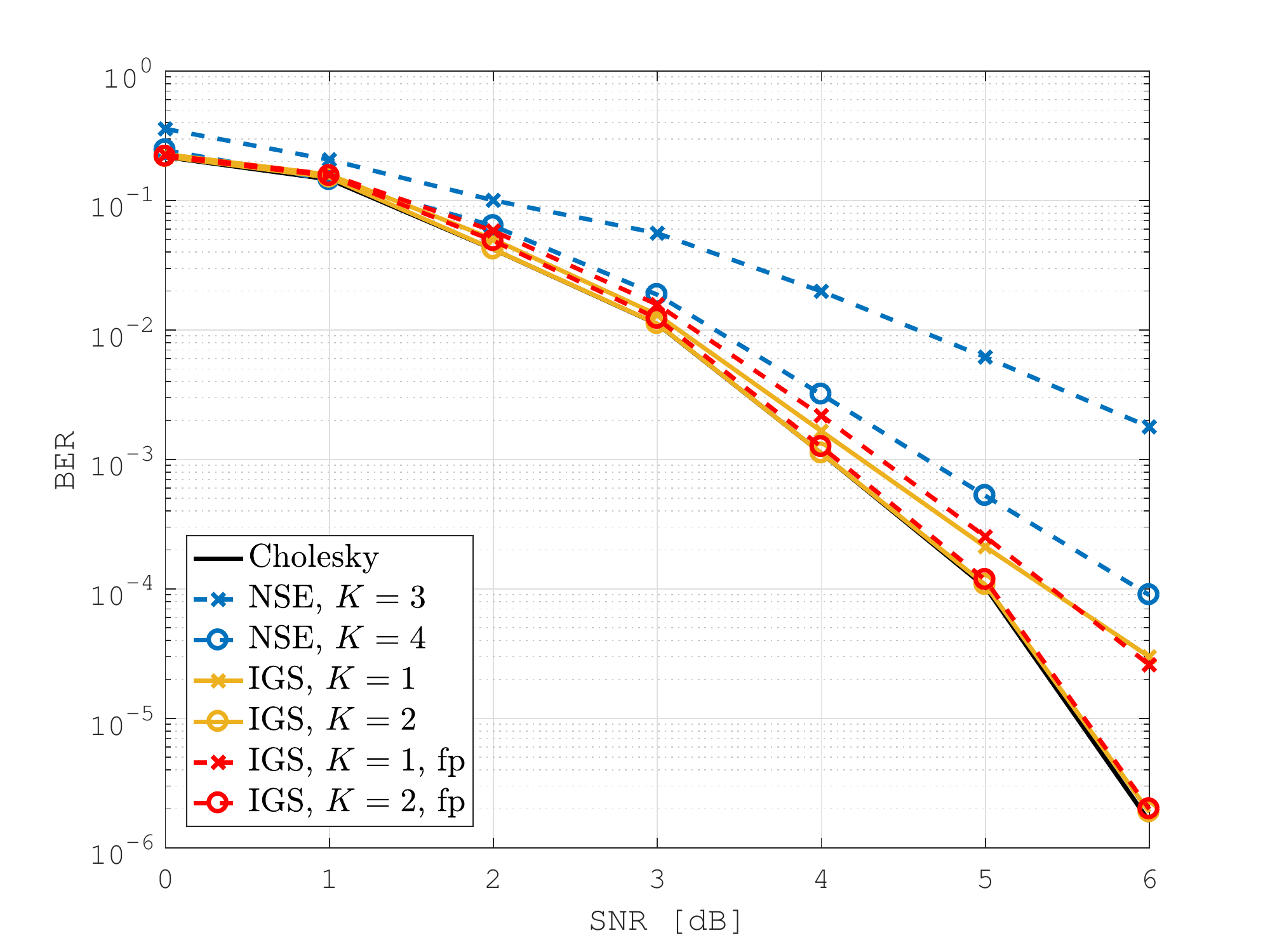}}
  \caption{Fixed-point simulation result for proposed quantization scheme.}
  \label{fig:fpsimu} 
\end{figure*}
\subsection{Fixed-Point Error-Rate Performance}
In order to achieve near-optimal error-rate performance with lower hardware complexity, fixed-point arithmetic is used in this implementation. The associated word-lengths for the proposed architecture have been determined via numerical simulations. The parameters provided in the following refer to the real or imaginary part.

For PU, the channel matrix $\mathbf{H}$, the receive vector $\mathbf{y}$, and the noise variance $N_0$ are represented with $15$ bit; the output of RGM is also quantized to $15$ bit and it is then compressed to $9$ bit with the proposed data compression scheme; the output of MF is set to $15$ bit. For ISCU and GSMU, the input $\mathbf{D}$ and $\mathbf{E}$ are decompressed to $15$ bit at beginning. For SCU, the input and the output are set to $15$ bit and $12$ bit, respectively. For LCU, its input is represented with $12$ bit and its output is quantized to $10$ bit. Here all multiplications are mapped onto DSP48 slices. The MAC registers are set to $22$ bit. Each LUT in the reciprocal module has $1024$ addresses and $15$ bit outputs, implemented by a single B-RAM.

Fig. \ref{fig:fpsimu} compares BER performance of the proposed fixed-point scheme and the floating-point algorithms for $128 \times 8$ and $64 \times 8$ systems. One can observe that the degradation introduced by the fixed-point scheme is negligible,
compared to the floating-point result. Specifically, the implementation
loss is less than $0.05$ dB SNR at $0.1\%$ BER.
\subsection{FPGA Implementation Results and Comparison}
\begin{figure}
\centering
\includegraphics[width=3.4in]{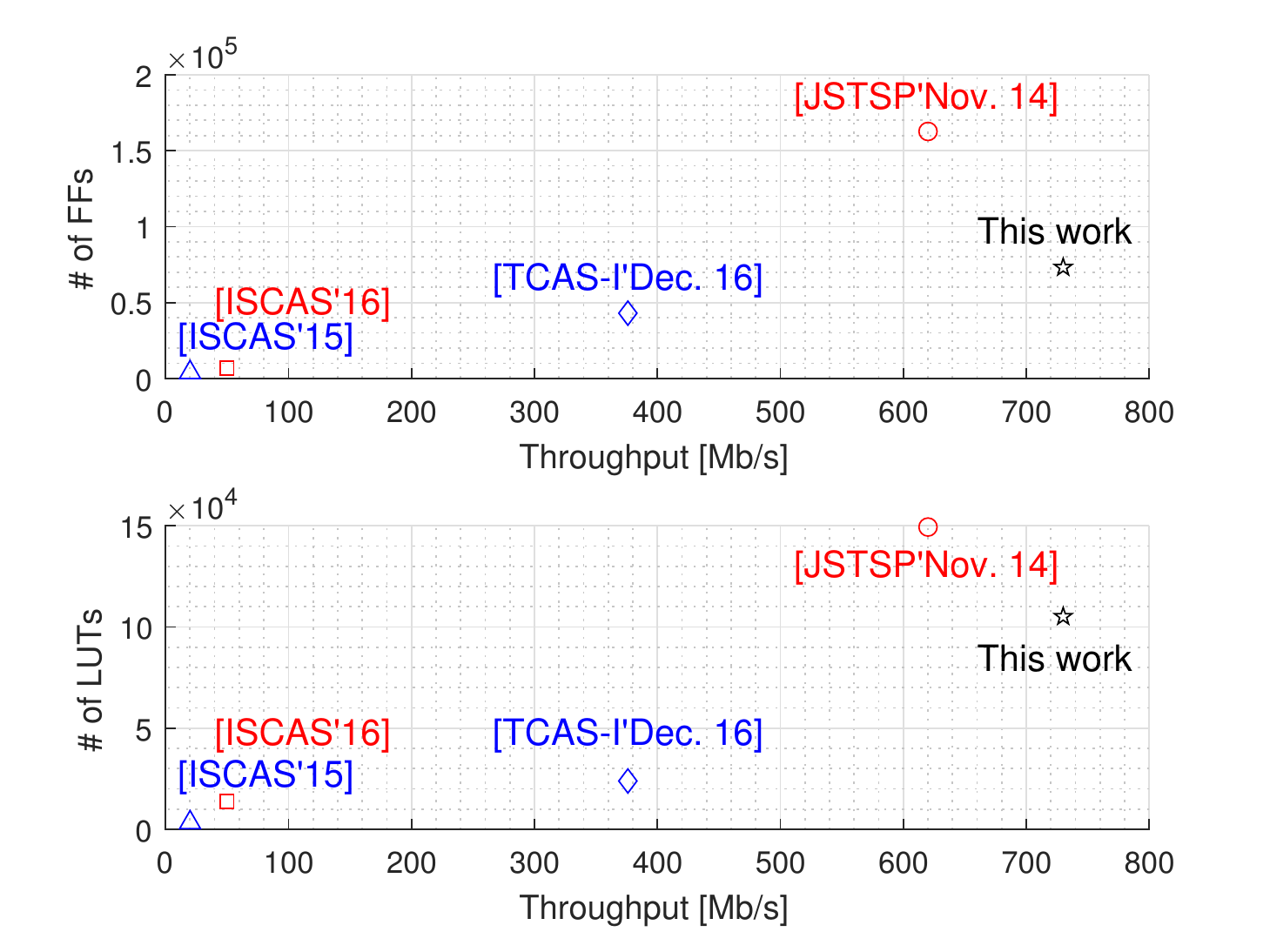}
\caption{Implementation comparison of state-of-the-art massive MIMO detectors.}
\label{fig:imple}
\end{figure}
Table \ref{tbl:results} provides the key implementation results of our GS-based soft-output massive MIMO detector and compares it with the post-place-and-route results of other state-of-the-art massive MIMO detectors \cite{Wu2014nse,Yin2015iscas,Castaneda2016iscas, Wu2016cd}. In addition, Fig. \ref{fig:imple} compares these detectors in terms of throughput and hardware overhead.

As discussed in Section \ref{subs:simu}, the proposed IGS algorithm for massive MIMO uplink should be compared to other NSE-based algorithms with less number of iterations for the sake of fairness. Since the detectors presented in \cite{Wu2014nse,Yin2015iscas,Wu2016cd} are with $K=3$ iterations and a $2$-term NSE is used in our algorithm, the proposed IGS detector is implemented with only $K=1$ iteration.
As shown in Table \ref{tbl:results}, our IGS detector has a much higher throughput ($732$ Mb/s) than the other detectors. One can also achieve a higher clock frequencies via increasing the number of pipeline stages, if the hardware complexity is affordable. Here we introduce the ratios of throughput/LUTs and throughput/FFs to measure the hardware efficiency of detectors. It is clear that our IGS detector achieves the highest hardware efficiency in terms of throughput/FFs, and the second highest hardware efficiency in terms of throughput/LUTs.

Furthermore, Fig. \ref{fig:ber_128}, \ref{fig:ber_64}, and \ref{fig:fpsimu} show that IGS with $K=1$ iteration outperforms the NSE-based one with $K=3$ in the case of different antenna configurations. In addition, for large system loading factors $N_t/N_r$ (e.g., $16/64$ and $8/64$), the conventional NSE-based detection algorithm performs much worse than the proposed GS-based one, even results in convergence issues.

Table \ref{tbl:thro} reports the throughput of IGS detector with different number of iterations. It indicates that the proposed IGS detector with $K=2,3$ can still achieve higher throughputs than the other aforementioned detectors.
\begin{table}[htbp]
\tabcolsep 2mm
\renewcommand{\arraystretch}{1.2}
\footnotesize
\centering
\caption{Throughput for $K$ iterations with $64$-QAM, $128$ BS antennas, and $8$ users.}
\label{tbl:thro}
\begin{tabular}{l||l|l|l}
\Xhline{1.0pt}

  & $K=1$  & $K=2$ & $K=3$\\ \hline
LUTs  & $105,135$ & $105,135$ & $105,135$\\
\hline
FFs  & $73,130$ & $73,130$ & $73,130$\\
\hline
Latency [clks] & $202$ & $219$ & $236$\\
\hline
Maximum clock freq. [MHz] & $308$ & $308$ & $308$\\
\hline
Throughput [Mb/s] & $732$ & $675$ & $626$\\
\Xhline{1.0pt}
\end{tabular}
\end{table}
\section{Conclusion}\label{sec:con}
In this paper, we have proposed an efficient hardware architecture of the GS-based soft-output detection algorithm for massive MIMO systems. The proposed iterative algorithm employs a novel initial solution based on NSE to accelerate convergence.
The proposed detection algorithm has shown its advantage of both convergence rate and BER performance in various antenna configurations and propagation environments. The corresponding VLSI architecture has been also provided with low hardware complexity. To further reduce the word-length of the regularized Gram matrix, a hardware-efficient data compression/decompression scheme is proposed in this paper. Exploiting the Hermitian property of the $2$-term NSE, the proposed architecture for NSE computation has a much lower processing latency. The optimized architecture for the GS method reduces the latency of each iterative by half. The provided reference FPGA implementation results have shown that our GS-based massive MIMO detector achieves a medium throughput with a much lower hardware complexity and a better error-rate performance, compared to the conventional ones. Future work focuses on iterative data detection
and decoding in massive MIMO systems based on this work.



%

\ifCLASSOPTIONcaptionsoff
  \newpage
\fi




\balance

\bibliographystyle{IEEEtran}
\footnotesize
\bibliography{IEEEabrv,mybib}

\balance

%

\end{document}